\documentclass[aps,twocolumn,prb,amsmath,amssymb,superscriptaddress]{revtex4}
\usepackage{multirow}
\usepackage{graphicx,epsfig}
\usepackage{color}

\def \tr {\text{Tr}}

\def \beps {\widetilde{\epsilon}}
\def \bteps {\widetilde{\boldsymbol{\epsilon}}}
\def \up {\uparrow}
\def \down {\downarrow}
\def \tr {\text{Tr}}
\def \calh {{\cal{H}}}

\begin{document}
\title{Time dependent heat flow in interacting quantum conductors}
\author{Guillem Rossell\'o}
\affiliation{Institut f\"{u}r Theorie der Statistischen Physik, RWTH Aachen University,
D-52056 Aachen,  JARA - Future Information Technologies, Germany}
\affiliation{Institut de F\'{\i}sica Interdisciplin\`aria i de Sistemes Complexos
IFISC (CSIC-UIB), E-07122 Palma de Mallorca, Spain}
\author{Rosa L\'{o}pez}
\affiliation{Institut de F\'{\i}sica Interdisciplin\`aria i de Sistemes Complexos
IFISC (CSIC-UIB), E-07122 Palma de Mallorca, Spain}
\author{Jong Soo Lim}
 \affiliation{School of Physics, Korea Institute for Advanced Study, Seoul 130-722, Korea}
\begin{abstract}
We derive the frequency-resolved  heat current expression in the linear response regime for a setup comprised of reservoir, interacting central site, and tunneling barrier under the action of a time dependent electrical signal. 
We exploit the frequency parity properties of response functions to obtain the heat current expression for interacting quantum conductors. 
Importantly, the corresponding heat formula, valid for arbitrary AC frequencies, can describe photon-assisted heat transport. 
In particular, we analyze the heat transfer for an interacting multilevel conductor (a carbon nanotube quantum dot) coupled to a single reservoir. 
We show that the electrothermal admittance can reverse its sign by properly tunning the AC frequency. 
\end{abstract}

\maketitle

\section{Introduction}
 
The expeditious advance of circuit miniaturization requires the knowledge of heat flow in quantum devices in response to electrical fields. \cite{heatquantum} Due to this necessity thermoelectrical transport in quantum devices is a research field that has vigorously relaunched nowadays. \cite{thermoelectricity}
Nevertheless, so far, the activity  has focused on driving electrical currents by oscillatory forces  \cite{but93a,but93b} for charge qubit manipulation \cite{Gao13},  quantum emitter generation \cite{zho03,gus09} or quantum tomography purposes. \cite{Glattli14}  Low frequency measurements of the electrical admittance  in quantum $RC$ circuits provide
 information about the spectroscopic ($C_\mathcal{G}$ quantum capacitance) and resistive ($R_\mathcal{G}$ relaxation resistance) properties of 
coherent conductors \cite{gabelli06,gabelli07,del11} in which $R_\mathcal{G}=h/2e^2$ takes an universal value. \cite{fu93,christien96a,christien96b,pretre96,buttiker96} Interactions, such as charging effects \cite{nigg06} or Kondo correlations \cite{minchul,mora}  do not alter the universality of $R_\mathcal{G}$, measured experimentally,  in quantum capacitors, \cite{gabelli06,gabelli07,del11,gabelli12,hashisaka12} carbon nanotubes, \cite{chorley12}  superconducting junctions, \cite{basset12} and quantum dots. \cite{frey12}

In contrast, time-resolved heat transport has been poorly investigated. \cite{heattime,dav13,dav14} 
Solely, stationary or  time-averaged heat flows have been analyzed in more detail. \cite{heatstationaryaveraged} 
The understanding of time dependent heat currents opens an avenue of creating circuit architectures where heat absorption or emission events are finely tunable via electrical and thermal time dependent signals. 
Recently, the linear response for the charge and heat fluxes to electrical and thermal AC signals was computed for a quantum capacitor showing that heat flows can be delayed or elapsed with respect to the AC pulse depending on the dot gate position. \cite{dav13}
Later on, Ludovico \textit{et al.} \cite{dav14} showed,  for a slow AC modulation and noninteracting conductors, that  time-dependent heat flow $\mathcal{J}_R(t)$ needs to  explicitly consider the energy stored and relaxed at the tunnel coupling region when a tunnel Hamiltonian description is employed. 
The goal that we face consists of the calculation of the heat flux in a completely different regime. 
We derive the time dependent heat flow for \textit{ interacting} and \textit{multi-orbital} conductors and \textit{arbitrary AC frequencies}. 
Since our heat formula holds for arbitrary high frequencies, it is able to describe photon-assisted tunneling processes occurring in the heat transport.  
Our only limitation is that we are restricted to the linear response regime, i.e., low AC driving amplitudes. 
We confine our interests to the Coulomb blockade regime and do not consider higher-order correlations like Kondo effect, 
we implicitly assume a temperature higher than the Kondo scale ($T_K$), i.e.,  $T  \gg T_K$. 
For our calculations, we employ the nonequilibrium (Keldysh) Green's function formalism \cite{jauho94} which allows us to include electron-electron interactions, charging effects, in a feasible way within the so-called Hartree-Fock regime. 

\begin{figure}[!t]
\begin{center}
\includegraphics[width=8 cm]{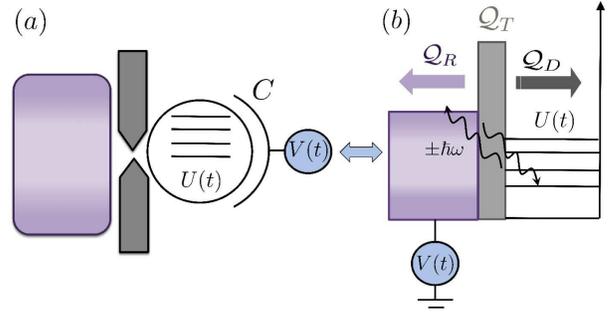}
\caption{Schematic of the multi-orbital interacting quantum capacitor. It is shown two situations where the AC voltage is applied either (a) to the dot or (b) to the reservoir. 
Both cases are equivalent by means of a gauge transformation. $U(t)$ denotes the time-dependent internal potential of the dot as response of the charge injected by the AC driving field $V(t)$. 
We model such response with a capacitance $C$.  We illustrate in (b) the induced photon-assisted emission and absorption processes (shown by wiggle lines) in the heat by the action of $V(t)$.  $\mathcal{Q}_R$, $\mathcal{Q}_D$, and $\mathcal{Q}_T$ are the energy change rates in time at the reservoir, quantum dot, and tunneling barriers.}
\label{figure1}
\end{center} 
\end{figure}
 
To address these issues, we focus on a prototypical  \textit{interacting multi-orbital quantum circuit}-- a quantum capacitor-- formed by a carbon nanotube quantum dot that is coupled to a single terminal being modulated by an AC voltage, as shown in Fig.~\ref{figure1}. 
Carbon nanotubes exhibit charging effects due to the formation of quantum dots inside the tube. \cite{bockrath97}  
The valley degrees of freedom, corresponding to the K and K' Dirac points in graphene in addition to the spin indices, lead to a four-fold energy level degeneracy. Such degeneracy can be lifted by the presence of an external magnetic field $B$. \cite{minot04}
Furthermore, the nanotube curvature yields a spin-orbit interaction resulting in split time-reversal dot level pairs. Therefore, carbon nanotube quantum dots act as multi-orbital interacting conductors where Coulomb interactions give rise to Coulomb blockade phenomena. 
Even more importantly,  carbon nanotube quantum dots have been demonstrated to be perfect platforms to investigate the frequency-resolved transport when they are embedded in electromagnetic cavities, \cite{chorley12} precisely the issue that our work addresses. 

In Sec.~\ref{model}, we describe the theoretical model for an interacting multi-orbital quantum capacitor and the Hartree-Fock decomposition of electron-electron interaction.
We derive the time-dependent heat flux for the reservoir in the linear response regime and obtain the electrothermal admittance.
For completeness we also discuss the electrical admittance. 
In Sec.~\ref{results}, we start by proving that our formalism can exhibit Coulomb blockade phenomena.
The observations for a single orbital quantum dot are briefly discussed and then results for a multi-orbital one are presented.
Here, we consider a four level degenerated carbon nanotube quantum dot which splits due to the presence of magnetic field and spin-orbit interaction. 
For this system, we show the $RC$ parameters and investigate the electrical and electrothermal admittances. 
Our findings are summarized in Sec. \ref{conclusions}.

\section{Time-dependent transport formulation}\label{model}

The purpose is to derive the linear response heat current for an interacting conductor coupled to a  reservoir that oscillates with AC voltage signal in the Coulomb blockade regime.  
We first propose a model Hamiltonian describing a multi-orbital quantum capacitor which is attached to a single reservoir. 
Using the Keldysh formalism for nonequilibrium Green's functions, we have the linear response heat flux at the reservoir subject to an oscillating electrical signal. Remarkably, the heat flux can be cast in terms of the conductor Green's function.
\subsection{Model of Hamiltonian}\label{model_of_hamiltonian}
The starting point of our derivation is the tunnel Hamiltonian description of a quantum capacitor. 
The quantum capacitor Hamiltonian $\mathcal{H}$  is split into three parts, namely the reservoir part ($\mathcal{H}_R$), the dot contribution ($\mathcal{H}_D$), and the tunneling term ($\mathcal{H}_T$), i.e.,
\begin{equation}
\mathcal{H}=\mathcal{H}_R+\mathcal{H}_D+\mathcal{H}_T\,.
\end{equation}
More concretely, the reservoir part reads
\begin{equation}
\mathcal{H}_R=\sum_{k,\sigma}\left[\epsilon_{k\sigma}-(\mu+eV(t))\right] c_{k\sigma}^\dagger 
c_{k\sigma}\,,
\end{equation}
where $\mu=E_F+eV_{dc}$ is the chemical potential, $e>0$ the unit charge, and $V(t)$ is the electrical voltage modulation [see Fig.~\ref{figure1} (b)].
As shown in Fig.~\ref{figure1}, notice that this description is equivalent to the situation where the AC voltage is applied to the dot since a gauge transformation connects both situations
and we thus employ two situations interchangeably.
Without loss of generality, we set $V_{dc}=0$. 
The operator $c_{k\sigma}^{\dagger} (c_{k\sigma})$ creates (annihilates) an electron with wavevector $k$ and spin $\sigma$ in the reservoir. 
For the dot contribution describing an interacting system with $n$ levels, we have
\begin{equation}\label{Hdot}
\mathcal{H}_D=\sum_{n,\sigma}\epsilon_{n\sigma}d_{n\sigma}^\dagger d_{n\sigma}+E_C\left[N_d+\mathcal{N}(t)\right]^2\,,
\end{equation}
where $d_{n\sigma}^{\dagger} (d_{n\sigma})$ corresponds to the creation (annihilation) operator for a dot electron in the $n$th level with spin $\sigma$, $\epsilon_{n\sigma}$ denotes the single-particle energies, 
and $E_C=e^2/2C$ is the electrostatic charging energy ($C$ is the dot geometrical capacitance). 
Here, the dot occupation operator reads $N_d=\sum_{n,\sigma} d_{n\sigma}^\dagger d_{n\sigma}$. 
Finally, the tunneling term hybridizes the reservoir and the dot subsystems according to
\begin{equation}
\mathcal{H}_T=\sum_{n,k,\sigma} t_{nk\sigma}\left(c_{k\sigma}^\dagger c_{n\sigma}+H.c\right).
\end{equation}
where $t_{nk\sigma}$ denotes the tunneling amplitude.  

When charge is injected in the dot by an external source $V(t)$, a polarization charge is created to keep the dot as a neutral charge object. In a simple electrostatic picture,  we model such polarization charge with a capacitance $C$ leading  to \cite{brun97} 
\begin{equation}
e\mathcal{N}_d(t)=C U(t)\,,
\end{equation} 
in which $U(t)$ is the internal potential of the dot giving rise to a time-dependent potential inside the dot. 
Then, the dot Hamiltonian in Eq. (\ref{Hdot}), up to linear order in $U(t)$, can be written as 
\begin{equation}
\mathcal{H}_{D}=\sum_{n,\sigma}\left[\epsilon_{n\sigma}+eU(t)\right] d_{n\sigma}^\dagger d_{n\sigma}+E_C N_{d}^2\,.
\end{equation}
Since $N_d^2 = \sum_{m,\sigma}\sum_{n,\sigma'} d_{m\sigma}^{\dag}d_{m\sigma}d_{n\sigma'}^{\dag}d_{n\sigma'}$ is a quartic operator, the Hamiltonian cannot be solved without introducing some proper approximation. 
\subsection{Hartree Fock approximation}
We thus perform the Hartree-Fock approximation in the dot Hamiltonian. For the Hartree approximation, $N_d^2$ is decoupled in the form
\begin{equation}
[N_d^2]_{\text{Hartree}} = 2\sum_{m,\sigma}\sum_{n,\sigma'} d_{m\sigma}^{\dag}d_{m\sigma}\langle d_{n\sigma'}^{\dag}d_{n\sigma'}\rangle\,,
\end{equation}
while for the Fock approximation one has
\begin{equation}
[N_d^2]_{\text{Fock}} = -2\sum_{m,\sigma}\sum_{n,\sigma'} d_{m\sigma}^{\dag}d_{n\sigma'}\langle d_{m\sigma}^{\dag}d_{n\sigma'}\rangle\,.
\end{equation}
Considering  Hartree and Fock approximations (HF) together, the dot Hamiltonian then becomes
\begin{equation}\label{Hdot_HF}
\mathcal{H}_D = \sum_{m,n,\sigma} \beps_{mn\sigma}(t) d_{m\sigma}^{\dagger}d_{n\sigma}\,,
\end{equation}
where
\begin{multline}
\beps_{mn\sigma}(t) = \delta_{m,n}\left(\epsilon_{m\sigma} + eU(t) + 2E_C\!\!\!\sum_{l,s (\ne m,\sigma)}\langle d_{ls}^\dagger d_{ls}\rangle\right)
\\ 
-2E_C\langle d_{m\sigma}^\dagger d_{n\sigma}\rangle ,
\label{renorlevel}
\end{multline}
with $\bar{\sigma} = \down/\up$ for $\sigma = \up/\down$.

\begin{figure}[!t]
\begin{center}
\includegraphics[width=8 cm, angle=-90]{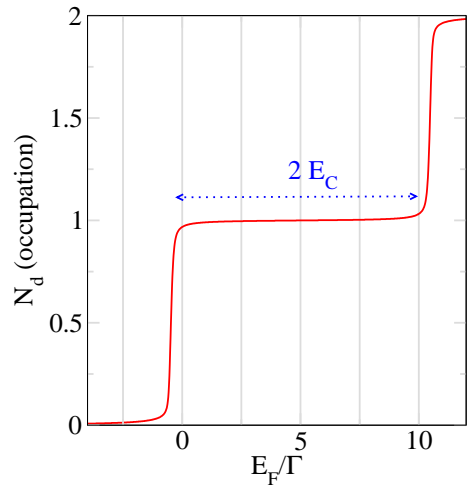}
\caption{Quantum dot occupation $N_{d}$ versus the Fermi energy $E_F$ for a single orbital dot in the Coulomb blockade regime. 
The dot's energy level is placed at $\epsilon_{d\sigma}=\epsilon_d+\sigma\Delta_Z$ with $\epsilon_d=0$ and $\Delta_Z=0.5\Gamma$. 
Rest of parameters: $2E_C=10\Gamma$ and $k_BT=0.04\Gamma$.}
\label{figure2}
\end{center} 
\end{figure}

An important observation is in order when Eq. (\ref{Hdot_HF}) is considered. For illustration, we assume that there is only a single orbital ($\epsilon_{m\sigma}=\epsilon_{d\sigma}$). 
We allow the presence of a small external magnetic field $\epsilon_{d\sigma}=\epsilon_{d}+\sigma \Delta_Z$ ($\Delta_Z$ the Zeeman energy) in order to break explicitly spin degeneracy, i.e.,  $N_{d\sigma}\neq N_{d\bar\sigma}$.
For a single orbital, since the exchange interaction is absent between electrons with same spins the Fock term disappears such that the dot's energy level can be simplified as 
\begin{equation}
\beps_{d\sigma}(t)=\epsilon_{d}+\sigma \Delta_Z  +eU(t) + 2E_C\langle d_{\bar{\sigma}}^\dagger d_{\bar{\sigma}}\rangle\,.
\end{equation}
We envision the level is occupied by the spin $\sigma$ electron with energy $\epsilon_{d\sigma} + eU(t)$.
When another electron with opposite spin $\bar{\sigma}$ enters the level, its energy is increased by
the charging energy $2E_C$. Therefore, as a function of the Fermi energy $E_F$ the dot occupation then does not change continuously, but shows plateaus and discontinuous jumps. 
These jumps are the clear evidence of Coulomb blockade phenomenon. 
We illustrate how HF approximation captures charging effects by considering  a single orbital quantum dot and plotting its total occupation $N_d$ when the Fermi energy varies in Fig.~\ref{figure2}. We consider the absence of AC signal.  
The dot occupation shows a plateau of $N_d\approx 1$ in the Coulomb gap region, i.e., $\epsilon_d\lesssim E_F\lesssim\epsilon_d+2E_C$. 
This result corroborates the fact that our HF description can reproduce charging effects and thus Coulomb blockade properly.
We emphasize that our results are not equivalent or expected when they are compared with previous findings using noninteracting models.

\subsection{Heat admittance}
Our main interest is focused on the rate of heat at the reservoir. For such purpose, we evaluate the time derivative of each component of the Hamiltonian which is given by
\begin{equation}
\mathcal{Q}_S=\frac{d}{dt}\langle \calh_S \rangle = \frac{i}{\hbar}\langle[\calh,\calh_S]\rangle + \left\langle\frac{\partial\calh_S}{\partial t}\right\rangle \,,
\label{dtH}
\end{equation}
where $S=\{R,D,T\}$ is referred to $R$ reservoir, $D$ dot, and $T$ tunneling term. 
Notice that in Eq.~\eqref{dtH} the last term is the power supplied by an external source and must be subtracted from the definition of the heat rate $\mathcal{Q}_{S}(t)$ (see below). 
Besides, since the Hamiltonian operator $\calh$ commutes with itself it is fulfilled 
\begin{equation}
\label{conservationE}
\frac{i}{\hbar}\langle [\calh,\calh] \rangle =\mathcal{Q}_R+ \mathcal{Q}_D+\mathcal{Q}_T=0\,.
\end{equation} 

As mentioned, our goal is to compute $\mathcal{Q}_R$. When Coulomb interaction is taken into account, we obtain the rate of heat at the reservoir, $\mathcal{Q}_R$, in terms of interacting dot Green's functions.  
Since the direct calculation of ${\mathcal{Q}_R}$ is cumbersome, our strategy is first to compute the energy change rates at the dot and tunnel barrier using Eq.~\eqref{conservationE}
\begin{equation}
\mathcal{Q}_R=-(\mathcal{Q}_D+\mathcal{Q}_T)\,.
\label{eq:QR}
\end{equation}
We recall that our calculation applies only for the linear response regime and thus we keep only the leading order contributions in the fields $V(t)$ and $U(t)$.  
We employ the nonequilibrium Keldysh Green's function formalism for the calculation of the energy change rates in time. \cite{jauho94} 
We start with the time derivative of $\calh_D$ first,
\begin{multline}
\frac{d}{dt}\langle \calh_D(t)\rangle = \sum_{m,n,\sigma}\left[\partial_t \beps_{mn\sigma}(t)\right]\langle d_{m\sigma}^{\dag}(t)d_{n\sigma}(t)\rangle
\\
+ \sum_{m,n,\sigma} \beps_{mn\sigma}(t)\partial_t\langle d_{m\sigma}^{\dag}(t)d_{n\sigma}(t)\rangle \,.
\end{multline}
The first term on the right hand side denotes the power developed by the AC source and does not contribute to the heat flow. 
The second term represents the energy flux going into the dot which can be written in terms of the dot Green's function
\begin{equation}\label{heatdot}
\mathcal{Q}_D(t) = -i{\rm Tr}\left[\bteps_{d\sigma}(t) \partial_t \textbf{G}_{d\sigma,d\sigma}^<(t,t)\right]\,.
\end{equation}
Hereafter, \textbf{bold face symbols} denote matrices such that their $(m,n)$ components are 
\begin{equation}
\begin{split}
[\bteps_{d\sigma}(t)]_{m,n}  &= \beps_{mn\sigma}(t)\,,  
\\ 
[\textbf{G}^{<}_{d\sigma , d\sigma}(t,t')]_{m,n} &= i\langle d_{n\sigma}^\dagger(t') d_{m\sigma}(t)\rangle\,.
\end{split}
\end{equation}
The trace means summations over energy levels and spin indices, i.e., $\tr = \sum_{m,n}\sum_{\sigma}$. Secondly,  from the definition of $\calh_T$, it can be shown that
\begin{equation}
\langle\calh_T\rangle = -i\tr\left[\textbf{G}_{d\sigma,d\sigma}^<(t,t)\textbf{t}_{\sigma} + \textbf{t}_{\sigma}^{\ast}\textbf{G}_{c\sigma,d\sigma}^<(t,t)\right],
\end{equation}
and thus the time variation of energy stored or relaxed at tunneling barrier $\mathcal{Q}_T(t)=\partial_t\langle\calh_T\rangle$ is given by
\begin{equation}
\label{heattunnel}
\mathcal{Q}_T(t)= -i\partial_{t}{\rm Tr}\left[ \textbf{t}^*_{\sigma} \textbf{G}^{<}_{c\sigma,d\sigma}(t,t)+
\textbf{G}^{<}_{d\sigma,c\sigma}(t,t) \textbf{t}_{\sigma}\right].
\end{equation}
Here, the reservoir-dot and dot-reservoir Green's functions are given, respectively, by
\begin{equation}
\begin{split}
[\textbf{G}^{<}_{c\sigma,d\sigma}(t,t')]_{k,n} &= i\langle d^\dagger _{n\sigma}(t' )c_{k\sigma}(t)\rangle \,, 
 \\
[\textbf{G}^{<}_{d\sigma,c\sigma}(t,t')]_{n,k} &= i\langle c_{k\sigma}^\dagger (t') d_{n\sigma}(t)\rangle \,,
\end{split}
\end{equation}
and $[\textbf{t}_{\sigma}^*]_{n,k} = t_{nk\sigma}^*$, and $\tr=\sum_{k,n,\sigma}$.

Using the standard technique of equation of motion, the reservoir-dot Green's function $\textbf{G}^{<}_{c\sigma,d\sigma}(t,t)$ can now be recast in terms of solely the dot Green's functions as follows, 
\begin{multline}
{\bf t}_{\sigma}^{*}\textbf{G}^{<}_{c\sigma,d\sigma}(t,t)
= \int \frac{dt_1}{\hbar}~[{\bf \Sigma}^{r}_{\sigma}(t,t_1)\textbf{G}^{<}_{d\sigma,d\sigma}(t_1,t)
\\
+{\bf \Sigma}^{<}_{\sigma}(t,t_1)\textbf{G}_{d\sigma,d\sigma}^a(t_1,t)] \,.
\label{leadtodot}
\end{multline}
Here, we have defined
\begin{equation}
\begin{split}
{\bf\Sigma}^{r}_{\sigma}(t_1,t_2) &= -\frac{i}{2}\hbar{\bf\Gamma}\delta(t_1-t_2) \,,
\\
{\bf\Sigma}^{<}_{\sigma}(t_1,t_2) &= i{\bf \Gamma}e^{-i\phi_{V}(t_1,t_2)}f(t_1-t_2)
\end{split}
\end{equation}
as the retarded and lesser self-energies \cite{nigg06} that contain the time-dependent fields. 
We have assumed  a momentum-independent tunneling amplitude $t_{nk\sigma} = t_{n\sigma}$ leading to hybridization strength given by $[{\bf\Gamma}]_{m,n} = 2\pi\rho_0 t_{m\sigma}^*t_{n\sigma}$ with $\rho_0=1/2D$  the density of states of the reservoir and $D$ the reservoir bandwidth. 
In such self-energies, it is defined
\begin{equation}
 \phi_V(t_1,t_2)=\int_{t_2}^{t_1}\frac{dt}{\hbar}~ eV(t)\,,
 \end{equation}
as the time-dependent phase due to the AC external potential and 
\begin{equation}
f(t_1-t_2)=\int \frac{d\epsilon}{2\pi} e^{-i \epsilon (t_1-t_2)/\hbar} f(\epsilon)\,,
\end{equation}
being $f(t_1-t_2)$ the Fermi-Dirac distribution function in the time domain.  Similarly, the dot-reservoir Green's function $\textbf{G}^{<}_{d\sigma,c\sigma}(t,t){\bf t}_{\sigma}$ is given by
\begin{multline}
\textbf{G}^{<}_{d\sigma,c\sigma}(t,t){\bf t}_{\sigma}
= \int \frac{dt_1}{\hbar}~[\textbf{G}^{r}_{d\sigma,d\sigma}(t,t_1){\bf \Sigma}^{<}_{\sigma}(t_1,t)
\\
+\textbf{G}_{d\sigma,d\sigma}^<(t,t_1){\bf \Sigma}^{a}_{\sigma}(t_1,t)] \,.
\label{dottolead}
\end{multline} 
Employing Eqs.~\eqref{leadtodot} and \eqref{dottolead}, we find $\mathcal{Q}_T$ is also expressed only in terms of the dot Green's functions.

Once the heat energy rates  for the dot and the tunneling parts ($\mathcal{Q}_D$, and $\mathcal{Q}_T$) are expressed in terms of the dot Green's function, the next step consists of computing such Green's functions in the presence of the time-dependent AC signal.  
For the dot's retarded/advanced Green's function, we have \cite{nigg06}
\begin{equation}
{\bf G}^{r/a}_{d\sigma,d\sigma}(t,t')=e^{-i\phi_U(t,t')} {\bf G}^{r/a,eq}_{d\sigma,d\sigma}(t-t'),
\end{equation}
where $\phi_U(t,t') = \int_{t'}^{t} (dt_1/\hbar)~ eU(t_1)$
and $\mathcal{{\bf G}}_{d\sigma,d\sigma}^{r/a,eq}(t_1-t_2)$ denotes the dot's retarded/advanced Green's function in equilibrium, i.e., in the absence of AC signal.
Finally, the dot's lesser Green's function is obtained by means of
\begin{multline}
{\bf G}^{<}_{d\sigma,d\sigma}(t,t') = 
\\
\int \frac{dt_1}{\hbar}~\int \frac{dt_2}{\hbar}~ {\bf G}^{r}_{d\sigma,d\sigma}(t,t_1) {\bf \Sigma}^{<}_{\sigma}(t_1,t_2) {\bf G}^{a}_{d\sigma,d\sigma}(t_2,t') \,.
\label{Gddlesser}
\end{multline}

Inserting Eqs.~\eqref{leadtodot}-\eqref{Gddlesser} into Eqs.~\eqref{heatdot} and \eqref{heattunnel}, 
it can be shown, after some cumbersome algebra, that to linear order in $V$ and $U$ the energy change rates in frequency domain become
\begin{subequations}\label{flows}
\begin{equation}
\mathcal{Q}_D(\omega)=i e\omega \tr\Bigr\{\int\frac{d\epsilon}{2\pi}~\bteps_{d\sigma}\mathcal{A}(\epsilon,\hbar\omega)[V(\hbar\omega)-U(\hbar\omega)]\Bigr\}\,,
\end{equation}
\begin{multline}
\mathcal{Q}_T(\omega)=ie\omega \tr \Bigr\{\int \frac{d\epsilon}{2\pi}~ (\hbar\omega+2\epsilon-2\bteps_{d\sigma}) \mathcal{A}(\epsilon,\hbar\omega) \\ 
\times [V(\hbar\omega)-U(\hbar\omega)]\Bigr\} \,,
\end{multline}
\end{subequations}
where $V(\hbar\omega)$, and $U(\hbar\omega)$ are the Fourier transforms of $V(t)$, and $U(t)$ respectively. Here, 
\begin{equation}
\mathcal{A}(\epsilon,\hbar\omega)=\Upsilon(\hbar\omega,\epsilon)\mathcal{F}(\epsilon,\hbar\omega)
\end{equation}
with  
\begin{equation}
\begin{split}
\Upsilon(\hbar\omega,\epsilon) &= \left[\mathcal{\textbf{G}}^{r,eq}_{d\sigma,d\sigma}(\epsilon+\hbar\omega) {\bf \Gamma} \mathcal{\textbf{G}}^{a,eq}_{d\sigma,d\sigma}(\epsilon)\right] \,,
\\
\mathcal{F}(\epsilon,\hbar\omega) &= \frac{f(\epsilon+\hbar\omega)-f(\epsilon)}{\hbar\omega} \,.
\end{split}
\end{equation} 
The dot's retarded/advanced Green's function in equilibrium and in frequency domain is given by
\begin{equation}
{\bf{G}}_{d\sigma,d\sigma}^{r/a}(\epsilon) = \frac{1}{\epsilon - \bteps_{d\sigma} \pm i{\bf\Gamma}/2} \,,
\label{Gddra}
\end{equation}
with 
\begin{multline}
[\bteps_{d\sigma}]_{m,n} = \delta_{m,n}\left(\epsilon_{m\sigma} + 2E_C\sum_{l,s (\ne m,\sigma)}\langle d_{ls}^\dagger d_{ls}\rangle\right)
\\
-2E_C\langle d_{m\sigma}^\dagger d_{n\sigma}\rangle \,.
\end{multline}
Although Eq.~\eqref{Gddra} has the same structure with its noninteracting retarded/advanced Green's function, the charging energy is included in the denominator such that it can describe the Coulomb blockade effect as mentioned above. 
To completely determine the dot Green's function, the diagonal and off-diagonal dot occupations need to be self-consistently calculated using
\begin{equation}
\langle d_{ns}^\dagger d_{ms}\rangle = \frac{1}{2\pi i}\int d\epsilon[{\bf{G}}_{d\sigma,d\sigma}^{<}(\epsilon)]_{n,m}\,.
\end{equation}

Finally, from Eqs.~\eqref{flows} and \eqref{eq:QR}, we obtain the expression for the energy change rate at the reservoir
\begin{multline}\label{Cflow}
\mathcal{Q}_R(\omega) = -(\mathcal{Q}_D+\mathcal{Q}_T)
=-ie\omega \tr\Big\{\int \frac{d\epsilon}{2\pi}  \\ 
\left(\hbar\omega+2\epsilon-\bteps_{d\sigma}\right)\mathcal{A}(\epsilon,\hbar\omega)
[V(\hbar\omega)-U(\hbar\omega)] \Big \} \,.
\end{multline}
Importantly, neither $\mathcal{Q}_R(\omega)$ nor  $\mathcal{Q}_D(\omega)$  have a well defined parity  when the AC frequency is reversed, 
and therefore, within linear response theory, these two magnitudes do not represent physical quantities [see Sec.~\ref{results}]. 
Based on these observations, the reservoir and dot frequency dependent heat current expressions must be thus modified to exhibit a proper parity property when the AC frequency is reversed.
We find the expressions
\begin{eqnarray}\label{rates}
\mathcal{J}_R= \mathcal{Q}_R(\omega)+\frac{1}{2}\mathcal{Q}_T(\omega), \quad
\mathcal{J}_D= \mathcal{Q}_D(\omega)+\frac{1}{2}\mathcal{Q}_T(\omega)\,.
\label{eq:JRJD}
\end{eqnarray}
satisfy the parity property. Remarkably, the choice of the factor $\frac{1}{2}\mathcal{Q}_T(\omega)$ is unique in order to ensure a well defined parity property in both $\mathcal{J}_R$, and $\mathcal{J}_D$.

Formally, these expressions agree with their noninteracting counterpart for the time dependent heat currents. \cite{dav14}
However, it should be noted that our theoretical analysis \textit{(i)} goes beyond low AC frequencies in contrast with Ref. [\onlinecite{dav14}] in which the heat rate was obtained up to second order in the AC frequency using the Floquet theory,  \textit{(ii) } includes the effect of Coulomb blockade, and\textit{ (iii)} is applicable to multi-orbital conductors in contrast to previous time heat formulations.  

Remarkably, our AC heat formula contains photon-assisted tunneling events only possible for sufficiently high AC  frequencies.
Note that a similar definition for the time-dependent heat, but applicable to a spin chain model (using a tight-binding model), was  proposed in Ref.~[\onlinecite{Wu}]. 
In such work, the heat flux connecting two sites, $i$ and $i\pm 1$, was incorporated to the general heat flow expression with a $1/2$ factor in close analogy to Eq.~\eqref{rates}.  
However, caution is needed when this comparison is made.  The results for a chain of sites are not immediately generalized  to our continuum model by just keeping the factor one-half in front of the tunneling energy flow.

After these considerations, the formulations for the the dot and reservoir frequency dependent heat currents  in the linear response regime read
\begin{multline}\label{heatdotdefi}
\mathcal{J}_D = -\mathcal{J}_R 
= ie\omega {\rm Tr} \Bigr\{\int \frac{d\epsilon}{2\pi}~ \left(\frac{\hbar\omega}{2}+\epsilon\right)\mathcal{A}(\epsilon,\hbar\omega)
\\ \times
[V(\hbar\omega)-U(\hbar\omega)]\Bigr\}\,.
\end{multline}
Importantly, this formula is the central finding of our work.

\subsection{Electrothermal admittance}

In the presence of a time-dependent driving force, it is quite general to characterize the transport using the concept of admittance. 
The complex electrothermal admittance is defined as
\begin{equation}\label{electrothermal}
\mathcal{M}(\omega)=\frac{\mathcal{J}_R(\omega)}{V(\hbar\omega)}.
\end{equation}
Notice that Eq.~\eqref{electrothermal} which can be obtained from Eq. (\ref{heatdotdefi})  contains the unknown function $U(\hbar\omega)$. 
Therefore, for a complete characterization of the complex electrothermal admittance, we need first to determine the internal potential  $U(\hbar\omega)$. 
For such purpose,  we note that the displacement current $\mathcal{I}_D$ can be featured by a capacitance $C$ in a simple model [here, we consider the situation shown in Fig.~\ref{figure1} (a)]
\begin{equation}
\mathcal{I}_D (\omega)= -i\omega CU(\hbar\omega). 
\end{equation}
Due to current conservation, the displacement current is equal to the tunneling current $\mathcal{I}_T$ \cite{nigg06} for a quantum capacitor
\begin{equation}
\mathcal{I}_T(\omega)=g(\omega) [V (\hbar\omega)-U(\hbar\omega)],
\end{equation} 
where [see Ref.~[\onlinecite{nigg06}] for its explicit derivation]
\begin{equation}
g(\omega) = ie^2\omega \tr\left\{\int \frac{d\epsilon}{2\pi} \mathcal{A}(\epsilon,\hbar\omega)\right\}.
\end{equation} 
The internal potential is obtained when we impose current conservation $\mathcal{I}_D=\mathcal{I}_T$, then $U(\hbar\omega)$ reads
\begin{equation}\label{Uinternal}
U(\hbar\omega)= \frac{g(\omega) V(\hbar\omega)}{-i\omega C + g(\omega)}.
\end{equation}
Inserting Eq. (\ref{Uinternal})  into Eq.~\eqref{heatdotdefi} completely characterizes the linear response of the heat current to a time dependent voltage, i.e., the electrothermal admittance 
\begin{equation}
\mathcal{M}(\omega)=m(\omega) \frac{-i\omega C}{[-i\omega C +g(\omega)]}\, ,
\label{eq:Momega}
\end{equation}
with 
\begin{equation}
m(\omega)= ie\omega\tr\left\{\int \frac{d\epsilon}{2\pi} \left(\frac{\hbar\omega}{2}+\epsilon\right)\mathcal{A}(\epsilon,\hbar\omega)\right\}. 
\label{eq:momega}
\end{equation}
Similarly, the thermoelectrical admittance is defined as
\begin{equation}
\mathcal{L}(\omega)=\frac{\mathcal{I}_{T}(\omega)}{T(\omega)},
\end{equation}
 with $T(\omega)$ being the Fourier transform of a time modulated temperature $T(t)$. Remarkably, the thermoelectrical and
the electrothermal admittances are reciprocally related due to the microreversibility  principle
\begin{equation}
\mathcal{M}(\omega)=T\mathcal{L}(\omega).
\end{equation}
Notice that microreversibility principle only holds at linear order in $V(t)$, and $U(t)$. 
Finally, a second order expansion in the AC frequency for $m(\omega)$
\begin{equation}\label{Msecond}
m(\omega)=-i\omega C_{\mathcal{M}} + \omega^2 C_{\mathcal{M}}^2R_{\mathcal{M}}\,,
\end{equation}
allows us to obtain the $RC$ electrothermal parameters. 
For comparison, we also calculate the electrical admittance defined by
\begin{equation}
\mathcal{G}(\omega) = \frac{\mathcal{I}_T}{V(\hbar\omega)}
= g(\omega) \frac{-i\omega C}{[-i\omega C +g(\omega)]}\,.
\label{eq:Gomega}
\end{equation}
The corresponding second order expansion of $g(\omega)$ in frequency
\begin{equation}\label{Gsecond}
g(\omega)=-i\omega C_{\mathcal{G}}+ \omega^2 C_{\mathcal{G}}^2 R_{\mathcal{G}},
\end{equation}
always yields positive $RC$ parameters, i.e., $\mathcal{I}_T$ is  always delayed with respect to $V(\omega)$. This is in clear contrast to the electrothermal admittance case in which both $C_{\mathcal{M}}$ and $R_{\mathcal{M}}$
can be either positive or negative. The heat flow response is elapsed or delayed with respect to the electrical signal depending on the 
system parameters. Note however that the product of both quantities, the $RC$ time, is kept always positive as expected. Similar results were obtained in Ref. [\onlinecite{dav13}]. 

\begin{figure}[!t]
\begin{center}
\includegraphics[width=8 cm, angle=-90]{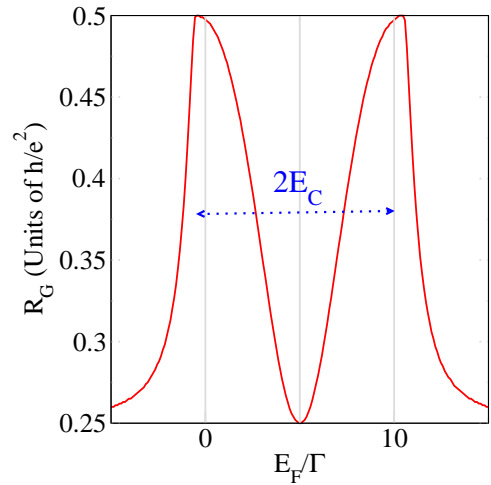}
\caption{Charge relaxation resistance for a single orbital quantum capacitor versus the Fermi energy. Parameters: $\epsilon_d=0$, $\Delta_Z=0.5\Gamma$, $2E_C=10\Gamma$,  and $k_BT=0.04\Gamma$. 
}
\label{figure3}
\end{center} 
\end{figure}


\begin{figure}[!t]
\begin{center}
\includegraphics[width=6 cm, angle=-90]{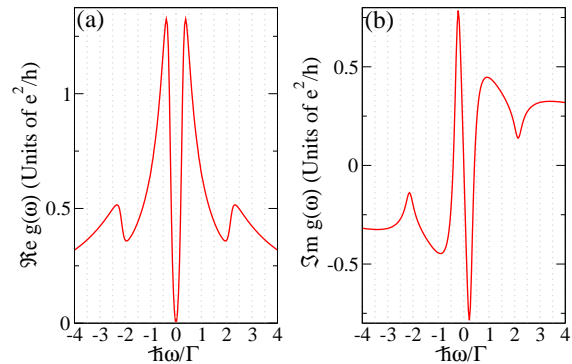}
\caption{(a) Real  and (b) imaginary parts of $g(\omega)$ for a single orbital quantum capacitor as a function of the AC frequency. 
Photon-assisted excitations occur at the transition rate energies $\pm\hbar\omega=|E_F-\epsilon_{d\sigma}|$ when $k_BT\ll \Gamma$.  
Parameters: $\epsilon_d=0$, $\Delta_Z=0.625\Gamma$, $2E_C=10\Gamma$, $E_F=0$, and $k_BT=0.04\Gamma$. }
\label{figure4}
\end{center} 
\end{figure}

\begin{figure}[!t]
\begin{center}
\includegraphics[width=6 cm, angle=-90]{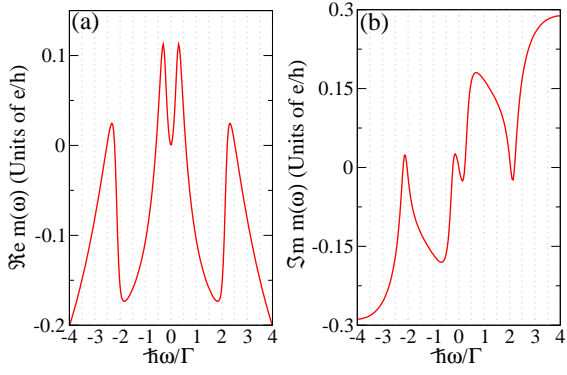}
\caption{(a) Real  and (b) imaginary parts of $m(\omega)$ for a single orbital quantum capacitor versus the AC frequency. 
Photon-assisted excitations occur at the resonant conditions.  Parameters: $\epsilon_d=0$, $\Delta_Z=0.625\Gamma$, $2E_C=10\Gamma$, $E_F=0$,  and $k_BT=0.04\Gamma$.}

\label{figure5}
\end{center} 
\end{figure}

\section{Results} \label{results}

Previously, we have derived the formal expressions for the reservoir heat flow and the corresponding electrothermal admittance. 
Now, we apply these formulas to the case of an interacting multi-orbital conductor. 
However, before addressing the multi-orbital case, for illustration purpose we first investigate a single orbital interacting quantum capacitor under the influence of an AC potential. 
Later on, we will analyze a prototype of multi-orbital conductor,  a carbon nanotube quantum dot attached to a single reservoir. 
In both cases, we show results for the electrical and electrothermal transport. 

\subsection{Single orbital quantum capacitor}
We consider an interacting quantum dot with just one orbital contacted to a single reservoir. 
The reservoir has an electrical potential that oscillates in time. 
In accordance with our previous theoretical considerations, the prefactor $g(\omega)$ of the electrical admittance $\mathcal{G}(\omega)$ can be expanded in powers of the AC frequency up to second order  as shown in Eq. (\ref{Gsecond}). 
A very well known result establishes that $R_{\mathcal{G}}=h/(2Qe^2)$ becomes universal being $Q$ the number of transport channels.  
Our example as shown in Fig.~\ref{figure3} considers the presence of a Zeeman field to  explicitly break spin degeneracy.  
$R_{\mathcal{G}}$ takes the value of $h/2e^2$ when just one of the two spin-resolved levels ($\epsilon_{d\uparrow/\downarrow}=E_F\pm \Delta_Z/2$) participates in transport, 
whereas it becomes $h/4e^2$ when the opposite spin channel also contributes.  
Clearly, the two main peaks observed in $R_{\mathcal{G}}$ are separated roughly by $2E_C$ the charging energy ($2E_C\gg \Delta_Z$). 
Our results for $R_\mathcal{G}$ indicate the presence of charging effects and the Coulomb blockade phenomenon. 

More importantly, our major interest resides in the behavior of the electrical and electrothermal admittances at arbitrary AC frequencies. 
Figure~\ref{figure4} shows the real [Fig.~\ref{figure4} (a)] and imaginary [Fig.~\ref{figure4} (b)] parts of the prefactor $g(\omega)$ in the expression for the electrical admittance $\mathcal{G}(\omega)$ [see Eq.~\eqref{eq:Gomega}].
We observe that $\Re e g(\omega) = \Re e g(-\omega)$ and $\Im m g(\omega) = -\Im m g(-\omega)$.
The factor $-i\omega C/(-i\omega C + g(\omega))$ in Eq.~\eqref{eq:Gomega} then does not change the parity property of $\mathcal{G}(\omega)$ with respect to $\omega$ such that for simplicity we consider $g(\omega)$. 
Excitations occur when the AC frequency matches with the resonant condition $\pm\hbar\omega\approx |\beps_{d\sigma}-E_F|$. 
We recall that $\beps_{d\sigma}$ is the spin-dependent dot energy level renormalized by electron-electron interactions according to
\begin{equation}
\beps_{d\sigma}=\epsilon_{d}+\sigma \Delta_Z  + 2E_C\langle d_{\bar{\sigma}}^\dagger d_{\bar{\sigma}}\rangle\,.
\end{equation}
For the parameters used in Fig.~\ref{figure4}, we obtain $\beps_{d\sigma}\approx 0.3\Gamma, \, 2.1\Gamma$ which agrees with the resonant behavior found in $g(\omega)$ with peaks at $\pm\hbar\omega\approx 0.3\Gamma$, and $ \pm\hbar\omega\approx  2.1\Gamma$.

Similar features are observed in the electrothermal admittance shown in Fig.~\ref{figure5}.   
Interestingly, the imaginary part of $m(\omega)$ for $\omega > 0$ takes either positive or negative values by tunning the AC frequency, which indicates that time heat current can be either delayed or elapsed with respect to the AC electrical time-dependent signal. 

Now, we discuss the parity property of the response functions $g(\omega)$ and $m(\omega)$ [thus $\mathcal{G}(\omega)$ and $\mathcal{M}(\omega)$].
We write $g(\omega)$ in the form
\begin{equation}
g(\omega) = \Re e g(\omega) + i\Im m g(\omega) \,,
\end{equation}
and express the real/imaginary part as
\begin{equation}
\Re e g(\omega) = \frac{1}{2} \left[g(\omega) + g^{\ast}(\omega)\right]
= \frac{1}{2}\int_{-\infty}^{\infty} dt~e^{i\omega t}\left[g(t) + g(-t)\right] \,,
\label{eq:Reg}
\end{equation}
\begin{equation}
\Im m g(\omega) = \frac{1}{2i} \left[g(\omega) - g^{\ast}(\omega)\right]
= \frac{1}{2i}\int_{-\infty}^{\infty} dt~e^{i\omega t}\left[g(t) - g(-t)\right] \,.
\label{eq:Img}
\end{equation}
Here, we used the fact that the response function $g(t)$ must be real to have a real expectation value for the current $\mathcal{I}_T(t)$.
From Eq.~\eqref{eq:Img}, it is quite easy to show that
\begin{equation}
\Im m g(-\omega)
= \frac{1}{2i}\int_{-\infty}^{\infty} dt~e^{i\omega t}\left[g(-t) - g(t)\right] \,,
\end{equation}
which implies $\Im m g(\omega) = -\Im m g(-\omega)$.
Using a similar line of reasoning, we can also prove that $\Re e g(\omega) = \Re e g(-\omega)$.
This argument also works for $m(\omega)$.
It is worthy to notice that this parity property comes from the fact that we have included the contribution due to the tunnel Hamiltonian in our definition for the heat flow as shown in Eq.~\eqref{eq:JRJD}.
Furthermore, this parity argument goes beyond the simple site partitioning scheme explained in Ref.~[\onlinecite{Wu}].

\subsection{Multi-orbital quantum capacitor }
We now investigate a single reservoir carbon nanotube quantum dot as an example of a \textit{multi-orbital interacting} conductor.
We regard the nanotube quantum dot as a localized single particle level described by two quantum numbers, the orbital quantum number $\tau$ associated to clockwise ($\tau=+1$) and anti-clockwise ($\tau=-1$)  semi-classical orbits along the nanowire circumference (related with the $K$-valley degeneracy in graphene) and the spin degree of freedom $\sigma$. 
In the presence of magnetic field along the nanotube axis, the dot energy level splits in the spin sector by the  Zeeman  field $\Delta_Z$ and in the orbital sector by an amount $\Delta_{\rm orb}$ that depends on the nanotube radius.  \cite{minot04} 
Besides, due to the nanowire curvature a non-negligible spin-orbit interaction is present, yielding a Kramers splitting of magnitude $\Delta_{\rm so}$. All together yields the following carbon nanotube dot energy level:  
\begin{equation}\label{resonancegeneral}
\epsilon_{d\sigma\tau}=\epsilon_d+\sigma\Delta_{Z}+\tau\Delta_{\rm orb}+ \sigma \tau \Delta_{\rm so}.
\end{equation} 
In the following, we show results that correspond to the realistic parameters: \cite{minot04} $\epsilon_{d}=0$, $\Delta_{Z}=0.625\Gamma$, $\Delta_{\rm orb}=5\Delta_Z$ (orbital magnetic moments can be $5-20$ times larger than its spin counterpart \cite{minot04}), and $\Delta_{\rm so}=0.5\Delta_Z$. 
The  charging energy is $2E_C=2\Gamma$ which lies in the strong interacting regime.  We restrict ourselves to the low temperature regime $k_BT=0.05\Gamma$. 
\begin{figure}[!h]
\begin{center}
\includegraphics[width= 6 cm, angle=-90]{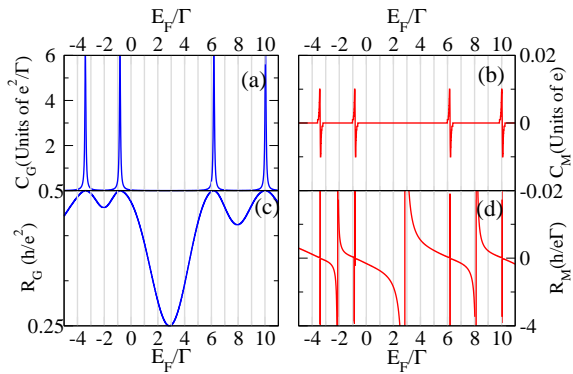}
\caption{(a) Electrical capacitance $C_\mathcal{G}$ and (c) electrical relaxation resistance $R_\mathcal{G}$ versus the Fermi energy $E_F$. 
(b) Electrothermal capacitance $C_\mathcal{M}$ and (d) electrothermal relaxation resistance $R_\mathcal{M}$ versus the Fermi energy $E_F$. 
Nanotube parameters: $\epsilon_{d}=0$, $\Delta_{Z}=0.625\Gamma$, $\Delta_{\rm orb}=5\Delta_Z$, $\Delta_{\rm so}=0.5\Delta_Z$, $2E_C = 2\Gamma$, and $k_BT=0.05\Gamma$.}
\label{figure6} 
\end{center} 
\end{figure}
    
First, we show our results for the electrical and electrothermal $RC$ parameters in Fig.~\ref{figure6}. In the low temperature limit ($k_BT\ll\Gamma$) the $RC$ parameters exhibit universal values. \cite{fu93,christien96a,christien96b}
We recall that in the low frequency regime the $RC$ parameters characterize the electrical and electrothermal admittances [see Eqs.~\eqref{Msecond} and \eqref{Gsecond}].  
$C_\mathcal{G}$ and $R_\mathcal{G}$  are displayed in Fig.~\ref{figure6} (a) and (c) when $E_F$ is varied. 
The electrical capacitance $C_\mathcal{G}$ shows oscillations which peaks at the positions located roughly at
\begin{eqnarray}\label{resonances}
 &&\epsilon_d-\Delta_Z-\Delta_{\rm orb}+\Delta_{\rm so} \approx -3.4\Gamma, \\ \nonumber
 &&\epsilon_d+\Delta_Z-\Delta_{\rm orb}-\Delta_{\rm so}+2E_C\approx -0.8\Gamma,  \\ \nonumber
 &&\epsilon_d-\Delta_Z+\Delta_{\rm orb}-\Delta_{\rm so}+2(2E_C)\approx 6.2\Gamma,  \\ \nonumber
 &&\epsilon_d+\Delta_Z+\Delta_{\rm orb}+\Delta_{\rm so}+3(2E_C)\approx 10\Gamma.
\end{eqnarray} 
 
 As expected,  at each nanotube level position, $R_\mathcal{G}$ takes the value of $h/2e^2$, while in the middle of two consecutive resonances  (when two resonances contribute to $R_\mathcal{G}$) it  diminishes to half of this value. 
Whereas $R_\mathcal{G}$ and $C_\mathcal{G}$ display always positive values, the electrothermal capacitance $C_{\mathcal{M}}$ and resistance $R_{\mathcal{M}}$ can become positive or negative when $E_F$ is tuned. 
This is shown in Fig.~\ref{figure6} (b) and (d). 
Indeed, $C_{\mathcal{M}}$ changes sign whenever the Fermi energy matches with any of the nanotube resonances. 
Heat current becomes delayed or elapsed with respect to the AC signal depending on the Fermi energy position. 
The electrothermal resistance $R_{\mathcal{M}}$ modifies its sign not only at the points when $E_F$ coincides with the nanotube resonances [see Eq.~\eqref{resonances}] but also when the electron-hole symmetry point occurs, just at the midpoint between two consecutive resonance points.  
The sign inversion in $R_{\mathcal{M}}$ happens when the derivative of the carbon nanotube quantum dot density of states vanishes. \cite{dav13} 
Remarkably, both $C_{\mathcal{M}}$ and $R_{\mathcal{M}}$ diverge around the resonance points, behavior that is washed out by enhancing temperature. 
(Similar results were obtained for the weak interacting limit, when $E_C\ll \Gamma$  see Ref. [\onlinecite{dav13}] for details).  
\begin{figure}[!h]
\begin{center}
\includegraphics[width=6 cm, angle=-90]{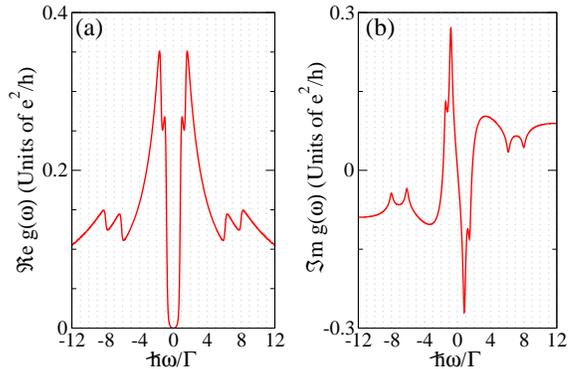}
\caption{ (a) Real  and (b) imaginary parts of $g(\omega)$ versus the AC frequency $\omega$. 
Parameters: $\epsilon_d=0$, $\Delta_Z=0.625\Gamma$, $\Delta_{\rm orb}=5\Delta_Z$, $\Delta_{\rm so}=0.5\Delta_Z$,  $2E_C=2\Gamma$, $E_F=0$, and $k_BT=0.05\Gamma$.}
\label{figure7}
\end{center} 
\end{figure}

We now discuss the case of arbitrary AC frequencies and analyze the prefactor $g(\omega)$ of the electrical admittance in Fig.~\ref{figure7}. 
The real and imaginary parts of $g(\omega)$ versus the AC frequency are depicted in Fig.~\ref{figure7} (a) and (b). 
As in the single orbital quantum capacitor, we observe that  $\Im m g(\omega)$ has odd parity with respect to $\omega$,
while $\Re e g(\omega)$ is an even function of $\omega$  as a consequence of being response functions of a real perturbating force. 
$\Im m g(\omega)$ accounts for the dissipative part of the electrical conduction with resonances roughly located at $\pm\hbar\omega \approx |E_F-\beps_{d\sigma\tau}|$. 
For the HF approximation, these resonances coincide with the dot level  positions that are renormalized by interactions according to Eq.~\eqref{renorlevel}.
These renormalized level positions (when $E_F=0$) are at $\beps_{d\sigma\tau}\approx -1.4\Gamma, \,-0.8\Gamma, \, 6.1\Gamma,\, 8.0\Gamma$ leading to the observed resonances in $g(\omega)$.  
The resonant behavior of $g(\omega)$  reflects the photon-assisted tunneling processes in which transport through the nanotube occurs by absorpting or emitting single photons. 
Furthermore, these resonances are broaden mainly by $\Gamma$ at very low temperatures. 
The $\Re e g(\omega)$ corresponds to the reactive part of the electrical conduction and has also a similar resonant structure. 
\begin{figure}[!h]
\begin{center}
\includegraphics[width=6 cm, angle=-90]{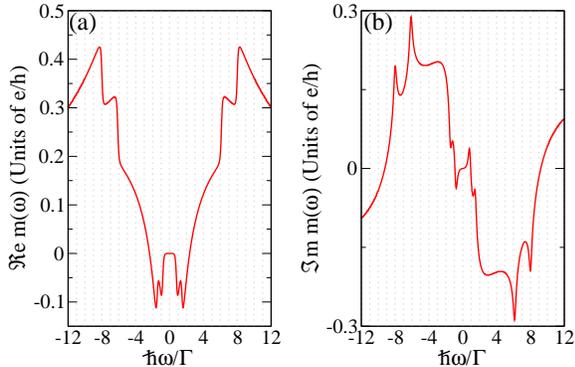}
\caption{(a) Real  and (b) imaginary parts of $m(\omega)$ versus the AC frequency $\omega$. 
Parameters: $\epsilon_d=0$, $\Delta_Z=0.625\Gamma$, $\Delta_{\rm orb}=5\Delta_Z$, $\Delta_{\rm so}=0.5\Delta_Z$,  $2E_C=2\Gamma$,  $E_F=0$, and $k_BT=0.05\Gamma$.}
\label{figure8}
\end{center} 
\end{figure}

We turn into the analysis of the electrothermal admittance $\mathcal{M}(\omega)$ which is given in Eq.~\eqref{eq:Momega}. 
As before, it is convenient to examine $m(\omega)$ defined in Eq.~\eqref{eq:momega} and we thus plot it as a function of the AC frequency in Fig.~\ref{figure8}. 
First, we observe $m(\omega)$ displays the same parity property as $g(\omega)$ which can be explained in the same way.
Furthermore, the real and imaginary parts of $m(\omega)$ are also characterized by features located at $\pm\hbar\omega\approx |E_F-\beps_{d\sigma\tau}|$. 
Remarkably, the observed peaks indicate single photon absorption and emission processes for the transport of heat. 
It is worthy to emphasize that our approach is able to capture such photon-assisted processes in contrast with other calculations restricted to  AC frequencies smaller or similar to the tunnel coupling $\Gamma$:
the other calculations assume the AC frequency is small such that only a second order expansion of $m(\omega)$ in powers of $\omega$ (characterized by the pairs $C_\mathcal{M}$, $R_\mathcal{M}$) can be justified. 
However, since we do not have such a restriction for the applied AC frequency, our results clearly exhibit photon-assisted  processes for the heat transport. 

Remarkably, $m(\omega)$ (either its real or imaginary part) takes positive or negative values depending on the AC frequency regime.
We observe $\Re e m(\omega)<0$ for low and moderate $\omega$, whereas for large AC frequencies its sign is reversed.   
We stress the importance of such result for the functionality of quantum circuits, in which one could manipulate the sign of the heat flow spectrum by properly tunning the AC frequency. 


\section{Conclusions}
\label{conclusions}
In closing, we have investigated the heat current spectrum in the linear response regime for an interacting conductor coupled to a single reservoir and modulated by an electrical AC signal. 
Our results, valid for arbitrary AC frequencies, show that the heat current expression for the reservoir needs to consider the heat stored or relaxed at the barrier. 
We illustrate our findings with two prototype of interacting conductors, namely,  a single orbital quantum dot and a multi-orbital conductor--a carbon nanotube quantum dot coupled to a single reservoir. 
We deal with the strong interacting limit where Coulomb blockade phenomena applies. 
We highlight that \textit{(i)} electrical and heat transport displays photon-assisted transport features, and \textit{(ii)} the electrothermal admittance can be positive or negative and the sign can be chosen by adjusting  properly the AC frequency. 
This is an important issue for engineering nanoelectronic circuits with optimal heat dissipation performances.

\section{Acknowledgements}
\label{acknowledgements}
We acknowledge J. Splettstoesser and D. S\'{a}nchez for critical reading and useful discussions. Work supported by MINECO Grant No. FIS2011-23526. We acknowledge financial support from the German Ministry of Innovation NRW.

\end{document}